# Cepheids in the Magellanic Clouds

D.L. Welch[1]

*Dept. of Physics and Astronomy, McMaster Univ., Hamilton, ON, L8S 4M1, Canada*

C. Alcock[2], D.P. Bennett[2], K.H. Cook

*Lawrence Livermore National Laboratory, Livermore, CA 94550*

R.A. Allsman, T.S. Axelrod, K.C. Freeman, B.A. Peterson, P.J. Quinn, A.W. Rodgers

*Mt. Stromlo and Siding Spring Observatories, Australian National Univ., Weston, ACT 2611, Australia*

K. Griest[2]

*Dept. of Physics, Univ. of California, San Diego, CA 92093*

S.L. Marshall

*Dept. of Physics, Univ. of California, Santa Barbara, CA 93106*

M.R. Pratt, C.W. Stubbs[2]

*Dept. of Astronomy, Univ. of Washington, Seattle, WA 98195*

and W. Sutherland (The MACHO Collaboration)

*Dept. of Physics, Univ. of Oxford, Oxford, OX1 3RH, U.K.*

**Abstract.** In the past few years, the Magellanic Clouds have been the targets for several major variable star surveys. The results of these surveys are now becoming available and it is clear that a Renaissance in LMC and SMC variable star research will result. In this review, I will describe the results of such surveys and review the questions that are likely to be answered by further work.

With respect to results, I will concentrate on LMC MACHO Project data, including beat Cepheids, discovery statistics, mode identification,

---

[1] MACHO Project Affiliate, The MACHO Project is a joint collaboration between the Lawrence Livermore National Laboratories (DOE), Mount Stromlo and Siding Spring Observatories (ANU), and the Center for Particle Astrophysics of the University of California (NSF)

[2] Center for Particle Astrophysics, University of California

> Fourier decomposition of lightcurves, and the differences between the LMC and galactic sample.

## 1. Introduction

The Magellanic Clouds continue to play a central role in our understanding of Cepheid variables. Long duration, wide-field surveys have now begun to yield large volumes of high-quality photometric data which will replace and extend the results of photographic surveys published over thirty years ago. Unlike these past surveys, the individual photometric measurements for each star will be available to all researchers. The impact of these surveys on our understanding of pulsating stars is not likely to be surpassed for decades to come.

## 2. Recent Work

Let me briefly mention work carried out since the review of Welch, Mateo & Olszewski (1993). The largest and most comprehensive published work is that of Smith *et al.* (1992) who surveyed 1.3 square degrees in the SMC for new variables using automated photographic photometry techniques. They discovered 78 new variables, many of them new Cepheids. Their mean B period-luminosity relation clearly shows two sequences of points which are interpreted as being due to Cepheids pulsating in the fundamental (F) and first overtone (1H) modes. They also find that 34% of the total sample are 1H pulsators. Sebo & Wood (1995) report on the detection of Cepheids in the rich, young clusters NGC 1850 and NGC 330. Storm *et al.* (1995) have undertaken Baade-Wesselink solutions for several SMC Cepheids. Poretti, Antonello & Mantegazza (1995) have produced the first results from a study of 21 short-period LMC Cepheids which compares the Fourier sequences in the Galaxy and the LMC in the neighbourhood of the long-period s-Cepheid sequence.

The two major intensive photometric surveys of LMC fields reported at this meeting are those of the EROS project by Beaulieu (1995) and the MACHO Project, described in more detail later in this paper, and by Cook (1995). The EROS project obtained some 15,000 images of a single 0.4 square degree field in both $B_J$ and $R_C$. They found 96 Cepheids in their field, 72 of which are new and found that 1H pulsators were 30% of their sample.

## 3. The MACHO Project

The work I would like to describe in somewhat more detail involves a portion of the MACHO Project photometry database. This survey project has been described by Alcock *et al.* (1992). Its principal goal is to detect microlensing events in order to estimate the mass spectrum of objects in the disk and halo. Due to the rarity of such events, it is necessary to observe large numbers of stars repeatedly, and in the process large numbers of variable stars are also detected and recorded. The 1.27m telescope at Mount Stromlo, Australia is currently dedicated full-time to this project. A camera built specifically for the project (Stubbs *et al.* 1993), has a field-of-view of 0.5 square degrees and images in a 'blue' (450-630nm) and 'red' (630-760nm) bandpass simultaneously.

To date the first 5500 frames, distributed over 22 fields and 400 nights, have been analysed. At present we have identified over 1500 Cepheids in these fields. Typical datasets contain between 100 and 320 two-colour observations.

### 3.1. Beat Cepheids

The passage of time is one of the few things in extragalactic research that can be measured with high precision. Beat Cepheids are stars which pulsate in two radial modes simultaneously. Only about 14 are known in our Galaxy, with CO Aur and EW Sct (Cuypers 1985; Figer *et al.* 1991) added to the list of Balona (1985). The ratio of the two periods can be determined with high precision and is expected to be dependent on metallicity. Beat Cepheids are difficult to discover, in practice, because of the large number of datapoints required to recognise and characterise them. Existing surveys for Cepheid variables in other galaxies have not detected beat Cepheids because they have lacked either the total number of observations required, the photometric precision, or both.

The MACHO photometric database was searched for stars whose phased lightcurves exhibited a photometric scatter in excess of that expected from photometric errors. A total of 45 beat Cepheids were discovered and are reported in Alcock *et al.* (1995). The mode identification was based on analysis of CLEAN'ed power spectra and position in the P-L and reddening-free P-L-C relations. MACHO Project lightcurves for a singly-periodic LMC Cepheid and an LMC beat Cepheid are shown in Figure 1. While the photometric calibration to a standard system is not complete, a number of conclusions can be drawn:

- Thirty are F/1H pulsators, fifteen are 1H/2H pulsators.

- The ratio of the shorter period ($P_S$) to the longer period ($P_L$) is near 0.7 and 0.8 for the F/1H and 1H/2H pulsators, respectively.

- The period ratios for LMC beat Cepheids are systematically higher than galactic beat Cepheids.

- The period ratio difference is presumably due to metallicity and is in the sense predicted by theory.

- Beat Cepheids seem to be found in bands across the instability strip (a result which must be confirmed by the photometric calibration).

- About 20% of the stars with fundamental periods shorter than 2.5 days are beat Cepheids.

- Several of the beat Cepheids were found in or near young clusters which will allow their evolutionary state to be determined.

- None of the previously reported beat Cepheid candidates common to our fields were confirmed.

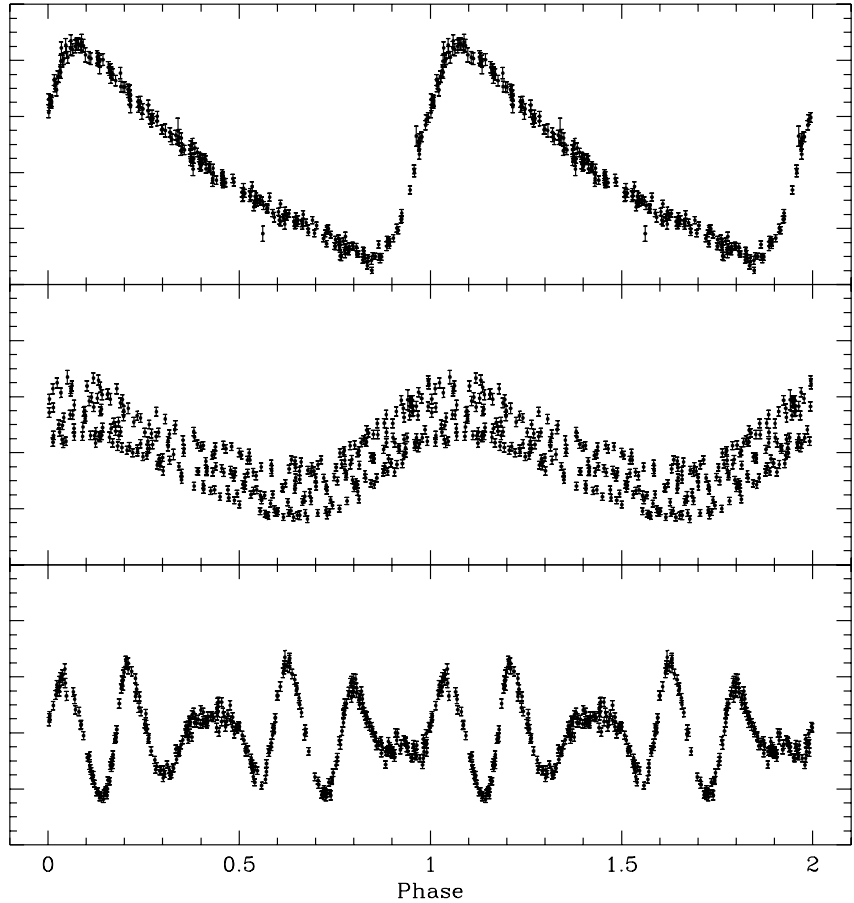

Figure 1. A $V_{\text{MACHO}}$ lightcurve of one of the 1500 Large Magellanic Cloud Cepheids in the MACHO Project photometry database is shown in the top panel. This fundamental pulsator has a period of 4.1 days. The middle panel illustrates the lightcurve of a so-called 'beat' Cepheid which is pulsating with two different periods simultaneously, when it is phased with on of the two periods (4.84 days). The bottom panel shows what the lightcurve would look like if followed continuously. The pattern in the lightcurve of this 'beat' Cepheid repeats every 24.2 days. There are a total of 249 observations for the normal Cepheid and 319 for the 'beat' Cepheid. Each point is plotted twice to reveal continuity. The vertical scale for each lightcurve is 1.0 mags.

Petersen & Christensen-Dalsgaard (1995) have already begun work on the theoretical implications of the LMC beat Cepheid data and hopefully will provide predictions for the expected period ratios for such stars with SMC-like metallicities.

### 3.2. Discovery Statistics

Of the 1500 Cepheids identified to date, 950 of these are F pulsators and 550 are 1H pulsators. We have cross-correlated our variable list the catalogue of Payne-Gaposchkin (1971) and find that new discoveries comprise 57% and 94% of the F and 1H samples, respectively. The lack of 1H pulsators in existing surveys is due to their relatively small photometric amplitudes. This survey is expected to be very nearly complete for Cepheids with amplitudes greater than 0.05 mag throughout the Cepheid period range.

### 3.3. Fourier Decomposition

The MACHO photometry database is a near-ideal sample for studying lightcurve behaviour. A systematic description of Cepheid lightcurve shapes in terms of Fourier series was first introduced by Simon & Lee (1981) as a means to compare observational light and radial velocity curves with the predictions of pulsation models. One drawback of the galactic Cepheid data sample is its observational inhomogeneity, despite heroic observing campaigns by numerous observers. Observations of LMC Cepheids allow a direct interpretation on the pulsation mode using the P-L and P-L-C relations - something not possible for their galactic counterparts.

We have undertaken Fourier decomposition of LMC Cepheid lightcurves. An example of the the run of the phase difference $\phi_{21}$ with $\log_{10} P$ is shown in Figure 2. In addition to allowing a comparison with theory, we are producing an algorithm by which the shape of a lightcurve of a given period and amplitude may be predicted. This will aid in the estimation of mean brightness of incompletely-sampled extragalactic Cepheid lightcurves.

### 3.4. Young Clusters Containing Cepheids

It is easy to get the impression that all young clusters containing Cepheids in the LMC and SMC have been surveyed. However, the work reviewed by Welch, Mateo & Olszewski (1993) covers only a small fraction of known young clusters and concentrates on those far from the bar of the LMC. We have cross-correlated positions of catalogued clusters with the existing set of 1500 Cepheid variables and find many new Cepheid-rich clusters. We plan to publish a list of these objects in the near future, with finder charts.

### 4. Future Avenues of Research

The mining of the MACHO Project photometric database has only begun. There are many avenues of research which we plan to explore. These include:

- We will search for evidence of singly-periodic 2H pulsators, and, if found, determine their lightcurve properties.

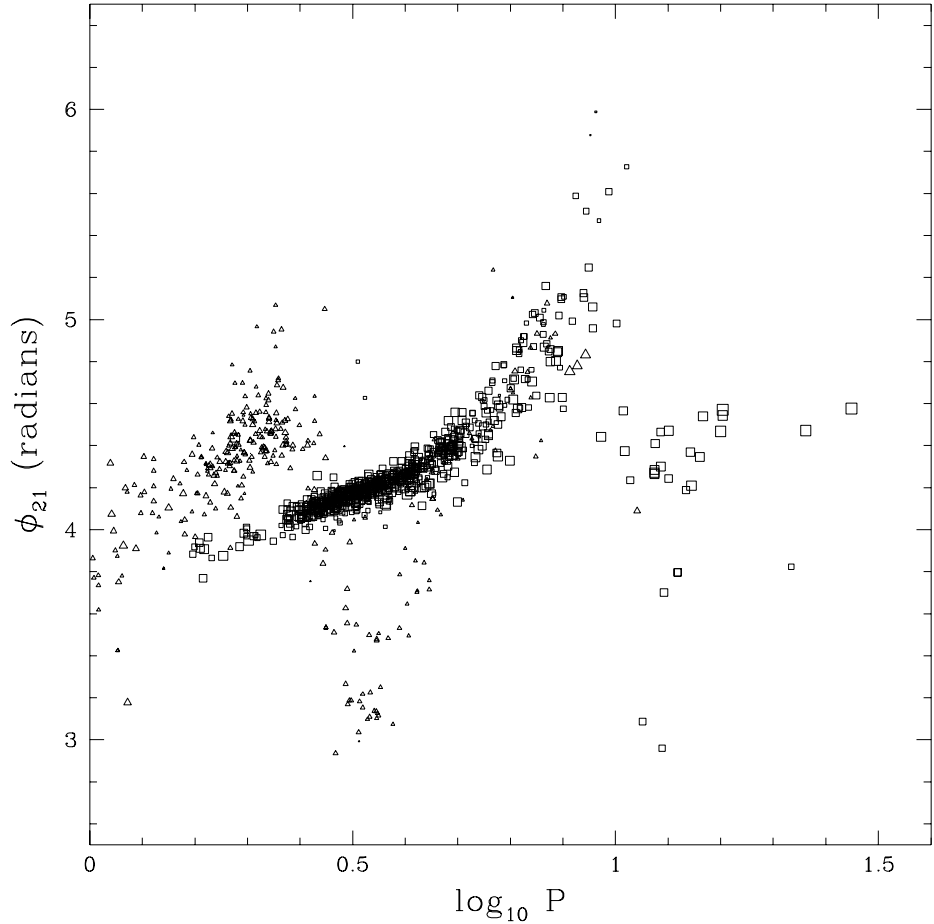

Figure 2. The Fourier parameter $\phi_{21}$ plotted as a function of $\log_{10} P$ for over 1500 Cepheids. This particular sample was selected by requiring the photometry lists to contain over 100 datapoints with fewer than 4 5-$\sigma$ outliers (which were excluded from the fit), a fitted RMS error smaller than 0.025 mag, and an error in $\phi_{21} < 0.15$ radians. The sample has not been carefully culled for contaminating objects such as foreground Type II Cepheids. The triangles and squares indicate an early (and imperfect!) classification of overtone and fundamental pulsation mode. The size of the symbols is proportional to the Fourier amplitude $R_1$. The sequence of long-period s-Cepheids is definitely present and the well-known resonance at 10 days are also clearly visible.

- When reductions for the six SMC fields observed by the MACHO Project become available, they will be searched for beat Cepheids and their period ratios will be compared with the LMC and galactic samples.

- We will make a systematic search for objects which change their lightcurve amplitude on a timescale of several years to determine whether or not objects like $\alpha$ UMi and V473 Cyg exist in the LMC.

- The remainder of the LMC MACHO fields will be searched and more recent data will included to improve existing period ratios.

- Clusters containing beat and singly-periodic Cepheids will be studied for evolutionary information.

- A catalogue of lightcurve parameters for the current sample will be compiled and released.

- The lightcurve data for all Cepheid variables will be compiled and released.

**Acknowledgments.** Work performed at LLNL is supported by the DOE under contract W7405-ENG-48. Work performed by the Center for Particle Astrophysics on the UC campuses is supported in part by the Office of Science and Technology Centers of NSF under cooperative agreement AST-8809616. Work performed at MSSSO is supported by the Bilateral Science and Technology Program of the Australian Department of Industry, Technology and Regional Development. DLW was a Natural Sciences and Engineering Research Council of Canada (NSERC) University Research Fellow during this work.